\documentclass[lettersize,journal]{IEEEtran}
\usepackage{amsmath,amsfonts}
\usepackage{algorithmic}
\usepackage{algorithm}
\usepackage{array}
\usepackage[caption=false,font=normalsize,labelfont=sf,textfont=sf]{subfig}
\usepackage{textcomp}
\usepackage{stfloats}
\usepackage{url}
\usepackage{verbatim}
\usepackage{graphicx}
\usepackage{cite}
\usepackage{cite}
\usepackage{graphicx}
\usepackage{url}
\usepackage{amsmath}
\usepackage{amssymb}
\usepackage{float}
\usepackage{booktabs} 
\usepackage{multirow}
\usepackage[colorlinks=true,linkcolor=blue]{hyperref}%
\usepackage{array}
\usepackage{color}
\usepackage{tabularx}
\usepackage{longtable}
\usepackage{tabu}
\usepackage{pifont}
\usepackage{amssymb}
\usepackage{ragged2e}
\usepackage{tabularx}
\usepackage{longtable}
\usepackage{booktabs}
\usepackage{rotating}
\newcommand\blfootnote[1]{%
  \begingroup
  \renewcommand\thefootnote{}\footnote{#1}%
  \addtocounter{footnote}{-1}%
  \endgroup
}

\begin{document}

\title{Compressing Human Body Video with Interactive Semantics: A Generative Approach}


\author{\IEEEauthorblockN{Bolin Chen\IEEEauthorrefmark{1},
Shanzhi Yin\IEEEauthorrefmark{1},
Hanwei Zhu\IEEEauthorrefmark{1},
Lingyu Zhu\IEEEauthorrefmark{1},
Zihan Zhang\IEEEauthorrefmark{1},\\
Jie Chen\IEEEauthorrefmark{2},
Ru-Ling Liao\IEEEauthorrefmark{2},
Shiqi Wang\IEEEauthorrefmark{1}\IEEEauthorrefmark{3} and
Yan Ye\IEEEauthorrefmark{2}}\\

\IEEEauthorblockA{\IEEEauthorrefmark{1} City University of Hong Kong} \hspace*{0.2in}  \IEEEauthorrefmark{2} Alibaba DAMO Academy \& Hupan Laboratory \\  \IEEEauthorrefmark{3} Shenzhen Research Institute, City University of Hong Kong}




\maketitle

\begin{abstract}
In this paper, we propose to compress human body video with interactive semantics, which can facilitate video coding to be interactive and controllable by manipulating semantic-level representations embedded in the coded bitstream. In particular, the proposed encoder employs a 3D human model to disentangle nonlinear dynamics and complex motion of human body signal into a series of configurable embeddings, which are controllably edited, compactly compressed, and efficiently transmitted. Moreover, the proposed decoder can evolve the mesh-based motion fields from these decoded semantics to realize the high-quality human body video reconstruction. Experimental results illustrate that the proposed framework can achieve promising compression performance for human body videos at ultra-low bitrate ranges compared with the state-of-the-art video coding standard Versatile Video Coding (VVC) and the latest generative compression schemes. Furthermore, the proposed framework enables interactive human body video coding without any additional pre- or post-manipulation processes, which is expected to shed light on metaverse-related digital human communication in the future.
\blfootnote{This work was supported in part by Shenzhen Science and Technology Program under Project JCYJ20220530140816037, in part by ITF GHP/044/21SZ and in part by Research Grants Council GRF-11200323.}

\end{abstract}

\begin{IEEEkeywords}
Interactive video coding, deep generative model, controllable embeddings, human body video.
\end{IEEEkeywords}

\section{Introduction}

Recent years have witnessed an explosive growth of human-oriented media contents in the emerging ``Short Video Era''. Although numerous progresses~\cite{wiegand2003overview,sullivan2012overview,bross2021overview,6020768} have been made to improve the compression and storage efficiency for human body video contents, there are still the following issues to be further addressed: 1) Is it possible to exploit statistical regularities of human body signals for compression efficiency improvement, thus realizing low-bandwidth human body video communication beyond traditional hybrid video coding frameworks? and 2) Is it possible to endower the encoded bitstream with semantic-level representations, thereby supporting direct interaction with the raw signals for metaverse-related digital human communication?

As for prior-based ultra-low bitrate video communication, it could be dated back to 1950's~\cite{7268565,1457470}. In particular, the strong prior information of a specific object, especially for talking faces, has been exploited in model-based analysis-synthesis codecs. However, limited by the past image processing abilities, the reconstruction quality was not satisfactory. Inspired by recent progress in video animation models~\cite{FOMM} based on Generative Adversarial Networks (GANs)~\cite{goodfellow2014generative}, the model-based coding technique can remedy its weaknesses in visual representation and reconstruction quality, thus advancing a novel generative compression paradigm~\cite{10533752,10647820} with promising performance. Taking generative face video compression as an example, the analysis model is employed to economically represent the input frames with compact facial representations (\textit{e.g.,} 2D landmarks~\cite{9455985}, 2D keypoints~\cite{ultralow}, 3D keypoints~\cite{wang2021Nvidia}, compact feature~\cite{CHEN2022DCC,chen2023csvt}, progressive tokens~\cite{chen2024beyond} and facial semantics~\cite{ifvc}), whilst the synthesis model is leveraged to achieve high-quality signal reconstruction via the powerful capabilities of deep generative models. 

Similarly to face signals, human body signals can also be compactly characterized based upon strong prior knowledge, thus achieving ultra-low bitrate communication within the analysis-synthesis-based generative compression framework. Specifically, Wang \textit{et al.}~\cite{10181664} and Yin \textit{et al.}~\cite{yin2024generative} exploited the compression potential of compact motion representations such as principal component analysis-based decomposing and multi-granularity temporal trajectory factorization. However, these implicit representations cannot describe well the non-linear dynamics and complex motion of human body signals, leaving the improvement room for compression efficiency and reconstruction quality. In addition, these implicit representations have no clear physical meanings in semantic level, thus cannot support the internal operations of motion features and the interactive functionalities for video communication.

\begin{figure*}[!t]
\centering
\centerline{\includegraphics[width=1\textwidth]{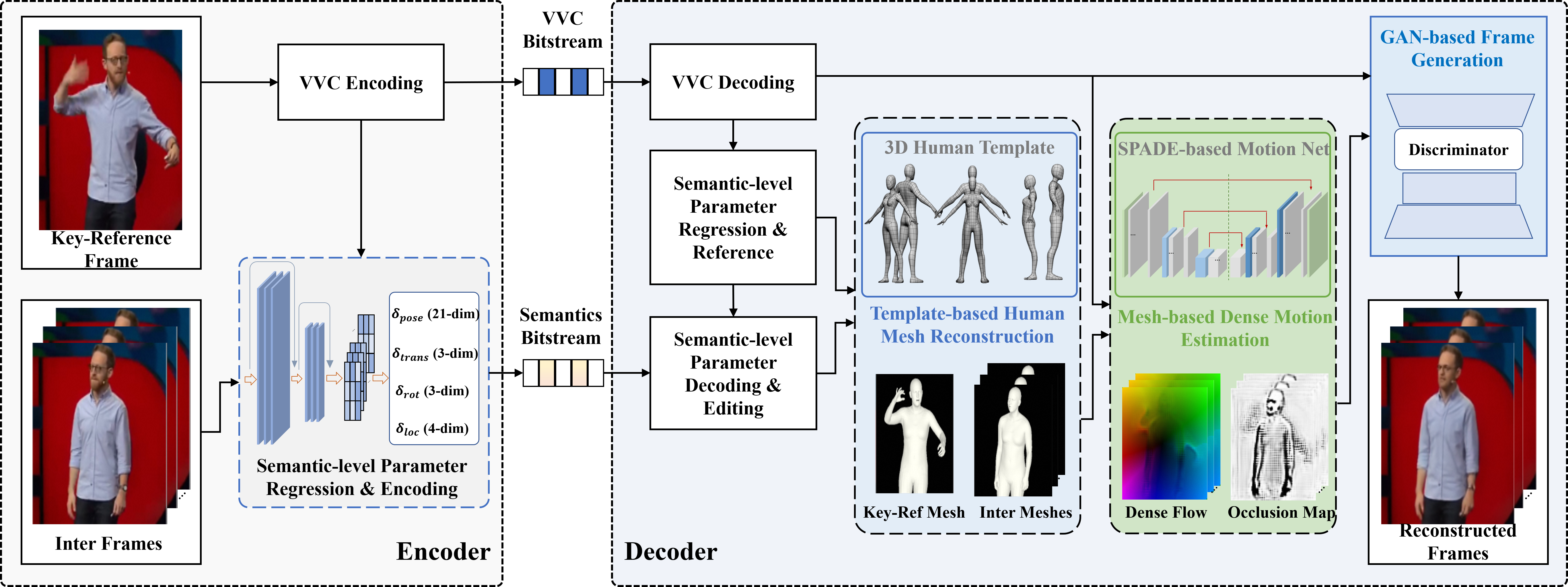}}
\caption{Overview of the proposed IHVC framework towards low-bandwidth and enhanced-interactivity human body video communication.}
\label{fig1}
\end{figure*}

To tackle these problems, we propose a novel Interactive Human Video Coding (IHVC) framework. To our best knowledge, it is the first generative compression framework which allows human body signals to be effectively encoded with ultra-compact and configurable human semantics. In this manner, the bitstream featured with these interactive human semantics can be feasibly manipulated, such that the head-pose and body-pose motions of human signal can be reconstructed towards personalized communication at the decoder side. Moreover, we design a mesh-based motion estimation scheme that could explicitly evolve these semantically meaningful representations into high-dimensional mesh representations with the assistance of 3D human model, thus facilitating dynamics awareness and quality improvement for reconstructing the desired decoded video. Experimental results illustrate that in comparisons with state-of-the-art algorithms, the proposed IHVC framework can attain a good balance among rate, visual quality and interactive functionalities, which is highly promising in ``short video'' applications and metaverse-related digital human communication.


\section{The Proposed Framework}
As illustrated in Fig.~\ref{fig1}, our proposed IHVC framework can realize low-bandwidth and enhanced-interactivity human body video communication. In particular, at the encoder side, the key-reference frame which represents the human body textures is compressed with the state-of-the-art VVC codec, which can provide texture reference for signal synthesis. Afterwards, the subsequent inter frames can be fed into a neural network-based parameter regressor for the characterization of compact and interactive semantics (\textit{e.g.,} 21-dim human pose parameters, 3-dim human translation parameters, 3-dim human rotation parameters and 4-dim human location parameters). Finally, these high-disentangled human semantics are further inter-predicted, quantized and entropy-coded into a transmitted bitstream.

When receiving the bitstream, the proposed decoder will perform the mesh editing/reconstruction and signal synthesis towards personalized interactions. First, key-reference frame is decoded via the VVC codec, and further projected into 3D human semantics. Besides, the compact semantics of inter frames are obtained by entropy decoding and compensation. Subsequently, the corresponding semantics of key-reference and inter frame are input into a preset 3D human template to reconstruct human body meshes, yielding the generation of the pixel-wise dense motion fields (\textit{i.e.,} dense flow and occlusion map). Herein, the reconstruction of 3D human body meshes can be flexibly edited by modifying different semantic-level representations, such that the head-pose and body-pose of 3D human meshes are varied towards personalized characterization. Finally, given the decoded key-reference frame and pixel-wise dense motion fields, the human body video can be high-quality reconstructed relying on the strong inference capability of deep generative models. 

To conclude, the proposed IHVC framework enjoys several desired advantages, including compact representations for ultra-low bitrate and semantically meaningful representations for enhanced interactivity. First, our scheme takes advantage of prior knowledge of human body signals, such that the 31-dimension semantic parameters are sufficient enough to characterize the nonlinear dynamics and complex motion of human body signals, thus facilitating the ultra-low bitrate human-oriented video communication. In addition, these highly-disentangled representations are interpretable with distinct semantic meanings in terms of body posture, head posture and camera location, which can be independently controllable for immersive interactivity and personalized reconstruction.

\subsection{Human Signal Parametrization/Modeling} 
Inspired by the progress in 3D human models, the input high-dimensional human body signal could be compactly characterized with semantic information and be reconstructed into the corresponding human mesh. Herein, we employ the pretrained 3D human model OSX~\cite{lin2023one} as our baseline module and optimize its parametrization/modeling processes for interactive compression functionalities.

\subsubsection{Semantic-level Parameter Regression \& Encoding}
Each input frame (\textit{i.e.,} the VVC reconstructed key-reference frame $\hat{K}$, or subsequent inter frames $I_{l} \left (1\le l \le n , l\in Z   \right ) $) is fed into the 3D human model OSX, thus obtaining a series of regressed semantic representations in terms of 3D separate body joint $\delta _{body}\in \mathbb{R}^{21\times3}$, body shape $\delta _{shape}\in \mathbb{R}^{10}$, 3D global translation $\delta _{trans}\in \mathbb{R}^{3}$, 3D global rotation $\delta _{rot}\in \mathbb{R}^{3}$ and a bounding box for body location $\delta _{loc}\in \mathbb{R}^{4}$. In total, the number of these OSX-regressed semantic parameters for each human body signal is 83, resulting in relatively high representation costs.

To improve compression efficiency with more economical representations, we further optimize the OSX-regressed semantic parameters. First, the human body frames from the same sequence share the matched human body shape $\delta _{shape}$, so these shape coefficients can be directly sourced from the reconstructed key-reference frame without signaling. In addition, by analyzing the physical meaning of separate body joint $\delta _{body}$, it can be found that only the last 7 joints are related to head posture and hand motions. Therefore, the last 21 dimensions from the 63-dimension vector are signaled as posture motion coefficients $\delta_{pose}\in \mathbb{R}^{7\times3}$ and the remaining first 42-dimension parameters can be obtained from the reconstructed key-reference frame. As for other semantics related with translation, rotation and location, they can be retained due to their compact representation and interactive function. To conclude, the final transmitted compact semantics $\delta_{sem}$ can be represented as, 
\begin{equation}
\label{eq1}
{
\delta_{sem}=\left \{  \delta _{pose}, \delta _{trans},\delta _{rot}, \delta _{loc} \right \},
}
\end{equation} 
where the total dimension of $\delta_{sem}$ signaled for each human body motion is 31. In addition, to achieve high-efficiency compression for these semantics, the inter-prediction operation between the current-frame semantics and the previously-reconstructed-frame semantics is further performed, and then the context-based arithmetic coding~\cite{TEUHOLA1978308,1096090} is applied to output the final bitstream. 

\subsubsection{Template-based Human Mesh Reconstruction}
When the decoder receives the semantics bitstream, the semantic coefficients of inter frames $\hat{\mathcal{\delta}}_{sem}^{I_{l}}$ can be reconstructed via the entropy decoding, inverse quantization and semantics compensation operations. Afterwards, these decoded semantic coefficients $ \hat{\mathcal{\delta}}_{sem}^{I_{l}}$ and other semantic parameters (\textit{i.e.,} the first 42 dimension of $\delta _{shape}^{\hat{K}}$ and $\delta _{body}^{\hat{K}}$) regressed from the reconstructed key-reference frame $\hat{K}$ will be jointly encapsulated and fed into the decoder of the OSX model $\mathbb{OSX}\left ( \cdot  \right ) $ to reconstruct 3D human mesh of inter frame $\mathcal M_{I_{l}}$ as follows,
\begin{equation}
\label{eq2}
{
\mathcal M_{I_{l}}=\mathbb{OSX}\left ( \hat{\mathcal{\delta}}_{sem}^{I_{l}}, \delta _{shape}^{\hat{K}},\delta _{body}^{\hat{K}} \right ).
}
\end{equation} 
As for the 3D human mesh of the reconstructed key-reference frame $\mathcal M_{\hat{K}}$, it can be directly generated via the OSX model without any parameter optimization. It should be mentioned that these 3D meshes $\mathcal M_{\hat{K}}$ and $\mathcal M_{I_{l}}$ will be further transformed to 2D mesh images (\textit{i.e.,}  $\mathcal M_{\hat{K}}^{2D}$ and $\mathcal M_{I_{l}}^{2D}$) for motion estimation.

\subsection{Mesh-based Dense Motion Estimation}
To better improve dynamics awareness for video reconstruction, we design a mesh-based motion estimation scheme that could evolve these transformed high-dimensional meshes into pixel-wise dense motion fields. In particular, the Spatially-Adaptive Normalization (SPADE) mechanism~\cite{park2019SPADE} (\textit{i.e.,} $\mathbb{S}_{\mathbb{PADE}}\left ( \cdot \right )$) is employed as the backbone network to predict these motion fields due to the fact that the SPADE network could well learn the reverse mapping from semantic maps $\mathcal M_{\hat{K}}^{2D}$ and $\mathcal M_{I_{l}}^{2D}$. Hence, the dense motion flow $\Gamma^{I_{l}}_{flow}$ and occlusion map $\Lambda ^{I_{l}}_{occlusion}$ can be obtained as the guidance of signal reconstruction as follows,
\begin{equation}
\label{eq3}
{
\Gamma^{I_{l}}_{flow}=P_{1}\left ( { \mathbb{S}_{\mathbb{PADE}}}  \left (  \hat{K}, \mathcal M_{\hat{K}}^{2D},\mathcal M_{I_{l}}^{2D}\right ) \right )  
},
\end{equation}
\begin{equation}
\label{eq4}
{
\Lambda ^{I_{l}}_{occlusion}=P_{2}\left ( {\mathbb{S}_{\mathbb{PADE}}}  \left (  \hat{K}, \mathcal M_{\hat{K}}^{2D},\mathcal M_{I_{l}}^{2D}\right ) \right )  
},
\end{equation} where $P_{1}\left ( \cdot \right )$ and $P_{2}\left ( \cdot \right )$ denote two different prediction functions integrating with the convolutional layer with different output channels for producing different guidance maps.

\subsection{GAN-based Human Frame Generation}
The strong inference capability of deep generative networks can well facilitate the realistic signal reconstruction within the generative compression paradigm. Herein, the GAN architecture is employed since it has obvious advantages in inference speed and application deployment compared with other generative models. More specifically, a feature warping strategy is performed to warp the decoded key-reference frame $\hat{K}$ with the dense motion flow  $\Gamma^{I_{l}}_{flow}$ in feature-level domain. Afterwards, to improve reconstruction fidelity, the occlusion map $\Lambda ^{I_{l}}_{occlusion}$ is used to indicate feature map regions with the corresponding confidences. The overall process can be described as,
\begin{equation}
\label{eq5}
{
{\hat{I}}_{l} =\Lambda ^{I_{l}}_{occlusion} \odot  f_{w}\left (\hat{K}, \Gamma^{I_{l}}_{flow} \right )
},
\end{equation} where $f_{w}$ and $\odot $ represent the back-warping operation and the Hadamard product, respectively. Finally, the generation results ${\hat{I}}_{l}$ are further fed into a discriminator module to approximate the distribution of the original signal ${I}_{l}$.

\subsection{Model Optimization}
Herein, we adopt the end-to-end strategy to jointly train the mesh-based dense motion estimation and GAN-based human frame generation modules, where the training losses are perceptual loss $\mathcal L_{per}$, adversarial loss $\mathcal L_{adv}$, texture loss $\mathcal L_{tex}$, pixel loss $\mathcal L_{pixel}$, DISTS loss $\mathcal L_{DISTS}$ as follows,
\begin{equation}
\label{eq6}
\begin{aligned}
\begin{array}{c}
\mathcal L_{total}=\lambda _{per}\mathcal L _{per}+\lambda _{adv}\mathcal L _{adv}+\lambda _{tex}\mathcal L _{text}
\\
+\lambda _{pixel}\mathcal L _{pixel}+\lambda _{DISTS}\mathcal L _{DISTS},
\end{array}
\end{aligned}
\end{equation} where $\lambda_{per}$, $\lambda _{adv}$, $\lambda _{texture}$, $\lambda _{pixel}$ and $\lambda _{DISTS}$ are set to 10, 1, 100, 20 and 20, respectively.

\begin{figure*}[!t]
\centering
\subfloat[Rate-DISTS]{\includegraphics[width=0.45 \textwidth,height=5.5cm]{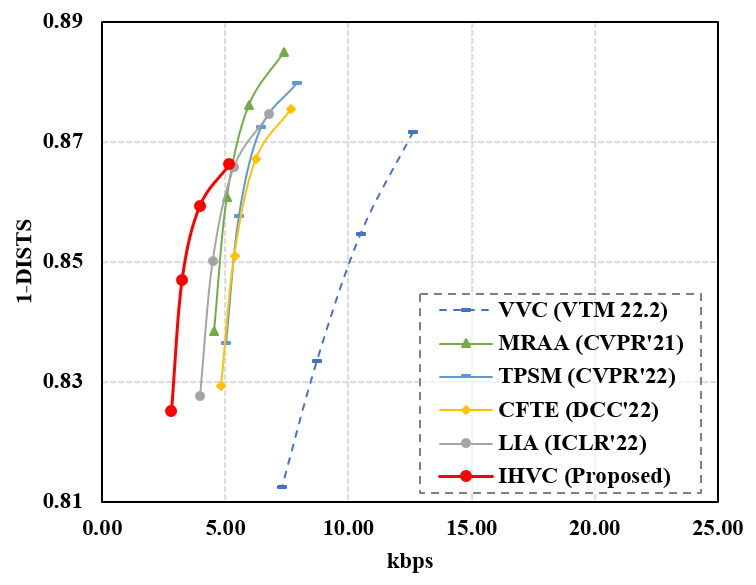}}
\subfloat[Rate-LPIPS]{\includegraphics[width=0.45 \textwidth,height=5.5cm]{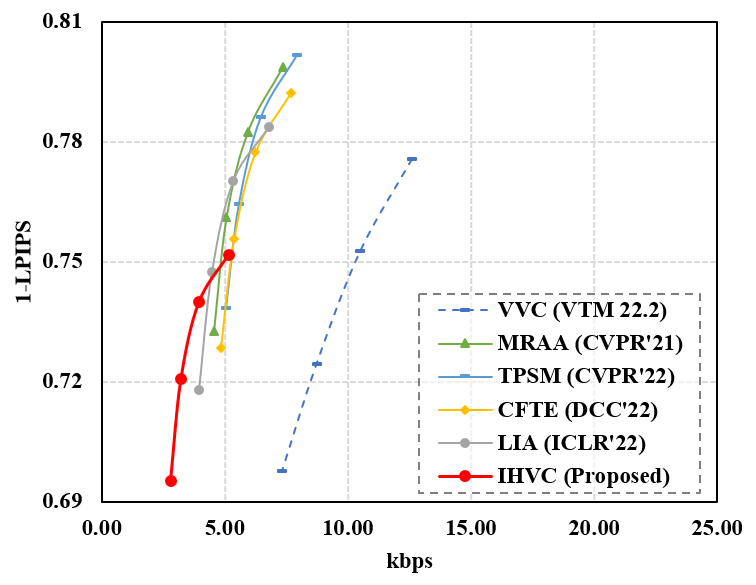}}
\caption{Overall RD performance comparisons with VVC~\cite{bross2021overview}, MRAA~\cite{mraa}, TPSM~\cite{tpsm}, CFTE~\cite{CHEN2022DCC} and LIA~\cite{lia} in terms of DISTS and LPIPS. }
\label{fig_RD}  
\end{figure*}

\begin{figure}[t]
\centering
\centerline{\includegraphics[width=0.4 \textwidth]{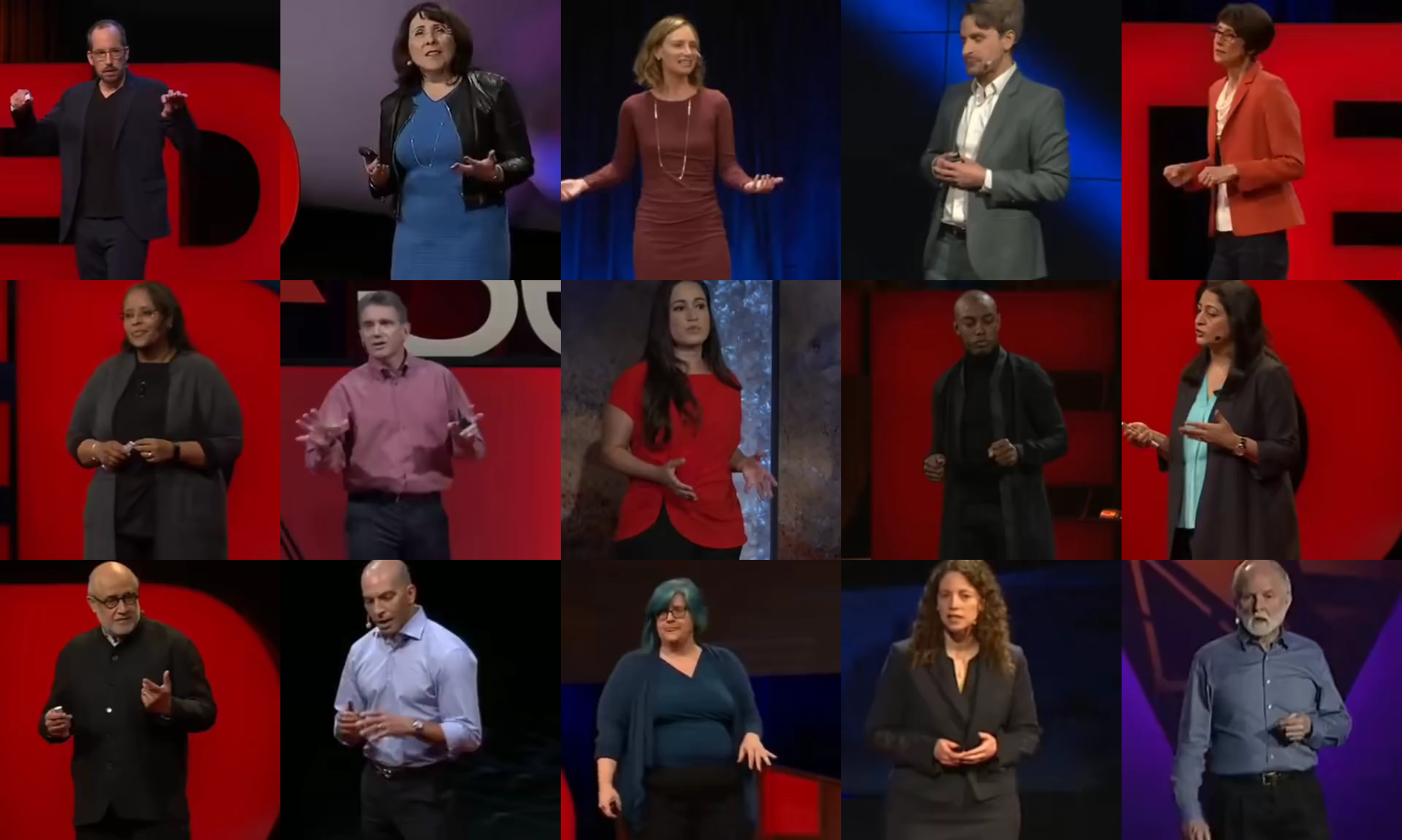}}
\caption{Illustration of 15 testing sequences selected and pre-processed from TED-Talk dataset~\cite{mraa}.} 
\vspace{-1em}
\label{fig2}
\end{figure}

\section{Experimental Results}
In this section, we first introduce the experimental settings, and then analyze experimental results in detail.

\subsection{Experimental Settings}
\textbf{Implementation Details.}
The proposed IHVC model is implemented with Pytorch libraries, and trained by the popular TED-Talk training dataset~\cite{mraa} with 300 epochs on 4 NVIDIA TESLA V100 GPUs. As for the TED-Talk dataset, it mainly contains 1,132 training videos and 128 testing videos at the resolution of 384$\times$384. During the model training phase, the Adam optimizer is set with the parameters $\beta _{1}$ = 0.5 and $\beta _{2}$= 0.999, as well as the learning rate is set at 0.0002 for model convergence. Besides, the synchronized BatchNorm initialization and the data repeating strategy are employed to improve the model robustness.

\textbf{Compared Algorithms.}
To demonstrate the effectiveness of the proposed IHVC, we compare the state-of-the-art video coding standard VVC~\cite{bross2021overview} and four representative generative human video compression algorithms, including MRAA~\cite{mraa}, TPSM~\cite{tpsm}, CFTE~\cite{CHEN2022DCC} and LIA~\cite{lia}. These testing algorithms are good performers in the TED-Talk testing dataset~\cite{mraa}. In particular, we select 15 testing sequences from this testing dataset, where each sequence contains 150 frames with the resolution of 384$\times$384. The specific settings of these compared algorithm are provided as follows,
\begin{itemize}
\item{\textbf{VVC Anchor:} The VTM-22.2 reference software is adopted and the coded mode is configured with the Low-Delay-Bidirectional (LDB). In addition, the Quantization Parameters (QPs) are set at 40/42/44/46 and the coded format for raw videos is YUV 4:2:0.}
\item{\textbf{Generative Codecs:} The key-reference frame of the raw video is first converted to the YUV 4:2:0 format and compressed via the VTM reference software 22.2 with the QPs of 27/32/37/42. As for the subsequent inter frames, they will be  represented into compact parameters, which can be further compressed via a context-adaptive arithmetic coder~\footnote{\href{https://github.com/nayuki/Reference-arithmetic-coding}{Context-adaptive Arithmetic Coder}}.}

\end{itemize}

\textbf{Evaluation Measures.} As reported in~\cite{10477607}, traditional pixel-level quality assessment methods like PSNR/SSIM may not be appropriate to evaluate generative video compression. 
Therefore, we adopt two learning-based perceptual quality measures (\textit{i.e.,} DISTS~\cite{dists} and LPIPS~\cite{lpips}) to evaluate the quality of these reconstructed human body videos.
For these two measures, smaller scores they have, better perceived quality they can perceive. As such, Fig. \ref{fig_RD} uses 1-DISTS and 1- LPIPS to present more coherent RD curves. In addition, we also employ the coding bitrate (\textit{i.e.,} kbps) to further evaluate the compression efficiency.

\begin{figure*}[!t]
\centering
\centerline{\includegraphics[width=1\textwidth]{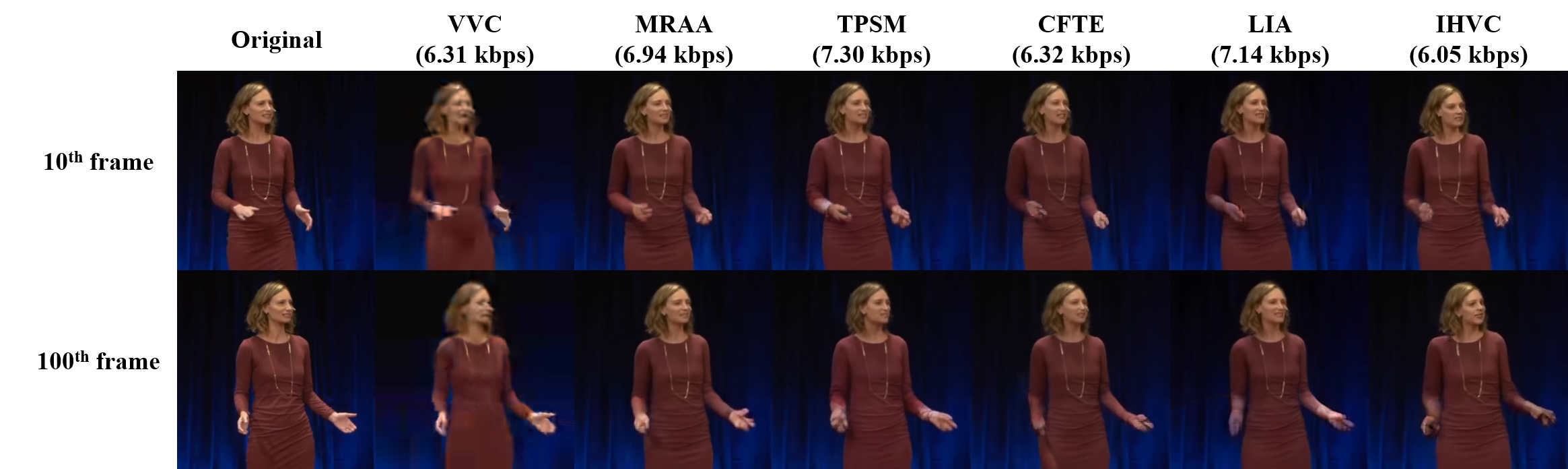}}
\caption{Visual quality comparisons among VVC~\cite{bross2021overview}, MRAA~\cite{mraa}, TPSM~\cite{tpsm}, CFTE~\cite{CHEN2022DCC}, LIA~\cite{lia} and IHVC (Proposed) at the similar bit rate. }
\label{fig3}
\end{figure*}

\begin{figure*}[!t]
\centering
\centerline{\includegraphics[width=0.8\textwidth,height=6.3cm]{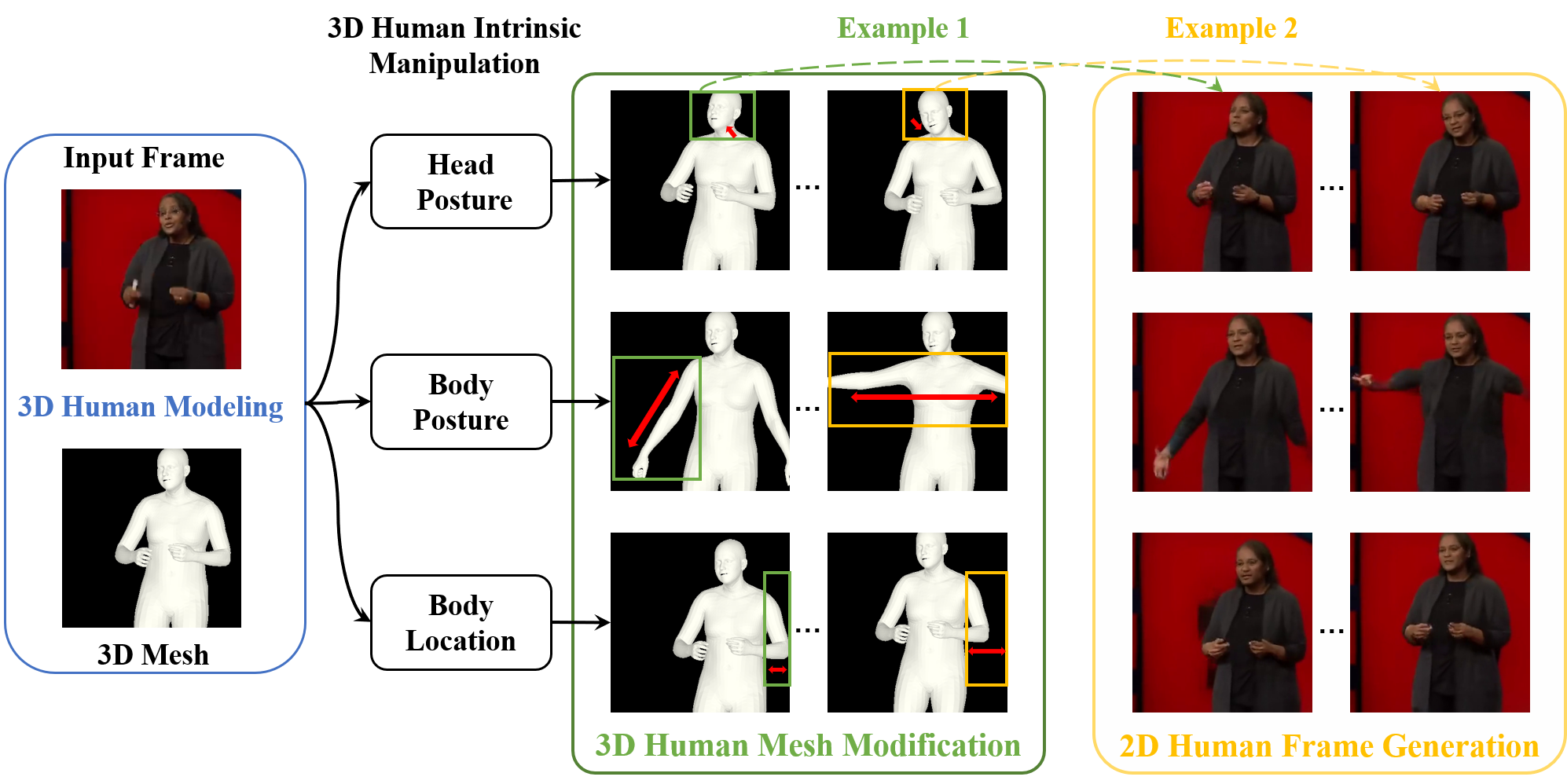}}
\caption{Illustration of interactive human body coding regarding different semantics on head posture, body posture and body location.}
\label{fig4}
\end{figure*}

\subsection{Performance Comparisons}

\textbf{Rate-Distortion Performance.} As shown in Fig. \ref{fig_RD}, our proposed IHVC scheme can achieve the optimal RD performance at ultra-low bitrate ranges in terms of DISTS and LPIPS measures compared with VVC and other generative compression algorithms. It should be noted that the (1-LPIPS) score of our IHVC scheme is relatively lower than those of other generative codecs, indicating an improvement room for the perceptual quality in future extended work.

\textbf{Subjective Quality.} Fig. \ref{fig3} shows the $10^{th}$ and $100^{th}$ frames of one particular sequences at similar bit rates among different compression schemes. Compared with the VVC codec, our proposed IHVC can reconstruct human video with higher fidelity and less artifacts. Besides, compared with other generation codecs, our proposed IHVC consumes lower bit rate whilst maintaining the reconstruction quality.

\textbf{Interactive Video Coding.} As illustrated in Fig. \ref{fig4}, our proposed IHVC framework can separately edit/control these transmitted semantics related with head posture, body posture and body location, thus ensuring great flexibility in personalized characterization and enhanced interactivity in metaverse-related human communication.

\section{Conclusions}
This paper proposes a novel IHVC framework to meet the exponential increase in the demand for low-bandwidth and enhanced-interactivity human body video compression. The fundamental principle of the proposed IHVC framework is enabling the high-dimensional signal representation with a series of compact semantics and allowing the direct manipulation of these semantics at the intrinsic representation level. Moreover, relying on the 3D human template and deep generative model, these manipulated semantics can be converted into high-dimensional meshes and estimate pixel-wise motion fields for the high-quality human body video reconstruction. Experimental results demonstrate that the proposed IHVC framework can achieve competitive RD performance and superior reconstruction quality for human body video compression compared with the VVC standard and other generative compression schemes. More importantly, it is also shown that the proposed IHVC framework can well support enhanced interactivity with head-pose and body-pose.

\bibliographystyle{IEEEbib}
\bibliography{main}
\end{document}